\documentclass{article}

\usepackage[english]{babel}

\usepackage[letterpaper,top=2cm,bottom=2cm,left=3cm,right=3cm,marginparwidth=1.75cm]{geometry}

\usepackage{amsmath}
\usepackage{graphicx}
\usepackage{authblk}
\usepackage[colorlinks=true, allcolors=blue]{hyperref}

\usepackage[authoryear,round]{natbib}

\title{5. Star-Planet Interactions: The Instrumentation Perspective }

\author[1]{H.~Korhonen\footnote{Corresponding author Heidi Korhonen, email: korhonen@mpia.de}}
\author[2,3]{P.~Figueira}
\author[4]{B. Montet}
\author[5]{S.P. J{\"a}rvinen}
\author[6,7]{H. Vedantham}
\author[8]{S. Narendranath}
\author[9]{P. Chaturvedi}

\affil[1]{Max-Planck-Institut für Astronomie, Königstuhl 17, D-69117 Heidelberg, Germany}
\affil[2]{Observatoire Astronomique de l’Universit\'{e} de Gen\`{e}ve, Chemin Pegasi 51b, 1290 Versoix, Switzerland}
\affil[3]{Instituto de Astrof\'{i}sica de Andaluc\'{i}a-CSIC, Glorieta de la Astronom\'{i}a s/n, 18008 Granada, Spain}
\affil[4]{School of Physics, University of New South Wales, Kensington 2052, Australia}
\affil[5]{Leibniz-Institut für Astrophysik Potsdam (AIP), An der Sternwarte 16, 14482 Potsdam, Germany}
\affil[6]{ASTRON, The Netherlands Institute for Radio Astronomy, Oude Hoogeveensedijk 4, 7991 PD, Dwingeloo, The Netherlands}
\affil[7]{Kapteyn Astronomical Institute, University of Groningen PO Box 800 9700 AV Groningen, The Netherlands}
\affil[8]{UR Rao Satellite Centre, Indian Space Research Organisation, Bengaluru, India}
\affil[9]{Department of Astronomy and Astrophysics, Tata Institute of Fundamental Research, Mumbai, India}

\date{}

\begin{document}
\maketitle

\begin{abstract}
\noindent Star-Planet Interactions (SPIs) produce observable phenomena across the electromagnetic spectrum, from X-ray and ultraviolet emission tracing magnetic activity and atmospheric escape to optical, infrared, and radio signatures probing stellar variability, planetary atmospheres, and magnetospheric interactions. This chapter reviews the current observational capabilities for SPI research, focusing on the instruments and facilities currently used, or technically capable of being used, for SPI studies. Particular emphasis is placed on the instrumental characteristics required to detect often weak and transient SPI signatures, as well as on the observational challenges associated with such measurements. The chapter concludes with an outlook on upcoming facilities that are expected to enhance our ability to detect, characterise, and understand SPI phenomena in the coming years.

\end{abstract}

\section{Introduction}

Star-planet interactions (SPIs) refer to interactions between a star and orbiting planets. Star-planet systems can exhibit four main types of interactions: magnetic, tidal, particle-based, and radiation-based. These different interactions that can lead to numerous phenomena, e.g., enhanced activity level of the host star, planetary aurorae, evaporation of the planetary atmosphere, and planetary migration. Their observation can be done over a wide range of wavelengths -- spanning from X-rays to radio -- and using numerous instruments and techniques. 

One of the most discussed signatures is enhanced stellar activity occurring phased with the planetary orbit. For example, chromospheric tracers such as Ca II H\&K, H$\alpha$, and ultraviolet (UV) emission have in some systems shown periodic modulation phased with the planet rather than stellar rotation \citep[see, e.g.,][]{Shkolnik+2003}, suggesting magnetic interactions between the stellar and planetary magnetospheres \citep[see, e.g.,][]{Cauley+2018}. However, these detections are often intermittent and difficult to distinguish from intrinsic stellar variability. In X-rays and UV, close-in planets may induce localised heating or magnetic reconnection events in the stellar corona, potentially producing excess emission or flaring activity \citep[see, e.g.,][]{Pillitteri+2010}. Some studies also report elevated X-ray luminosities in stars hosting hot Jupiters \citep[see, e.g.,][]{Poppenhaeger_Wolk2014}. Magnetised star-planet interactions may also generate coherent radio bursts through cyclotron maser emission detectable in radio frequencies \citep{hallinan2006,hallinan2008}. Table \ref{wavebands} gives a non-exhaustive overview of the different phenomena and signatures related to magnetic interactions.

Planetary atmospheric signatures can also reveal SPI processes. One of the important signatures are star induced escape of exoplanet atmospheres that can be observed in Ly$\alpha$ \citep{Vidal-Madjar+2003, Ehrenreich+2015}, He I 10830 {\AA} \citep{Spake+2018,Allart+2018}, or metal absorption lines \citep{Sreejith+2023}. Phase-dependent atmospheric escape \citep{Bourrier+2013} have been proposed as further evidence for magnetic star--planet interaction.

Another interesting line of investigation is to study potential auroral emission from exoplanets either using radio wavelengths \citep[see, e.g.,][]{vedantham2020} or H3$^{+}$ emission where prominent lines occur at 3--4 $\mu$m \citep[e.g.,][]{Richey-Yowell2025}. Detecting auroral emission would directly probe planetary magnetic fields, but unambiguous detections have so far remained elusive, partly because expected signals are weak and in radio occur at low  frequencies.

Massive close-in planets can influence stellar rotation, activity levels, and spin-orbit evolution through tidal dissipation. Observational signatures of these tidal interactions can, among others, include anomalously rapid stellar rotation \citep{Pont2009} and orbital decay \citep{Maciejewski+2016}. In extreme cases, tidal effects may also alter stellar pulsations \citep{Herrero+2011}.

In this chapter we discuss the different types of observations that can be used to study star-planet interactions and the astronomical instruments currently in use by the community to achieve this objective. We conclude with an outlook towards the future facilities technically capable of studying star-planet interactions.

\begin{table}
    \centering
    \caption{Overview of the magnetic star-planet interactions at different wavelength domains.}
    \label{wavebands}   
    \begin{tabular}{|l|l|l|}
    \hline
Wavelength & SPI Phenomena	& Typical Observational Signatures \\
\hline
Radio & Magnetic star-planet interaction; & Coherent radio bursts, cyclotron maser \\
 & auroral emission & emission, periodic low-frequency signals \\
\hline
Infrared & Atmospheric heating and escape; & He I 10830 {\AA} absorption, thermal phase,\\
 & auroral emission & variations, hot spots, H3$^{+}$ emission \\
\hline
Optical & Chromospheric and photospheric & Ca II H\&K enhancement, H$\alpha$ variability, \\
 &activity & activity synchronised with planetary orbit \\
\hline
Ultraviolet & Magnetic reconnection; & Enhanced UV line emission, flare activity, \\
 & atmospheric evaporation & excess chromospheric/coronal heating, \\
  &  & excess transit duration \\
\hline
X-rays & Coronal heating and & Enhanced X-ray luminosity, orbitally \\
 & magnetic interaction & modulated flares, coronal hot spots \\
\hline
\end{tabular}
\end{table}

\section{Optical and near-infrared spectroscopy from the ground}

Star-planet interactions in close-in planetary systems can produce observable spectroscopic signatures through their impact on the stellar atmosphere and, in some cases, provide indirect constraints on the atmospheric or magnetic properties of the planet. The most prominent signatures include enhanced emission in the cores of the Ca II H \& K lines, variability in the neutral helium triplet at 10830\,\AA, along with changes in many other spectral lines. In addition, the potential exoplanet aurorae could produce H3$^{+}$ emission detectable using near-infrared high-resolution spectrographs. Radial velocity observations can also provide insights in tidal interactions through detecting tidal orbital decay in close-in exoplanet systems. Long-term, high-precision spectroscopic monitoring can reveal gradual changes in orbital period and semi-major axis caused by tidal dissipation between the star and planet.

In the following, we review the instrumentation that has been used for studying SPI with ground-based spectroscopy.

\subsection{High resolution spectroscopy}

High-resolution spectroscopy is one of the primary tools for probing SPI because the expected signatures are often subtle, time-variable, and localized in spectrally resolved chromospheric or photospheric diagnostics. One of the most common methodologies used is the measurement of  emission in cores of chromospheric lines, notably in the Ca II H \& K lines (at 3968.469 \AA\ and 3933.663 \AA), which arises from enhanced chromospheric activity. When such variability is modulated with the planetary orbital period rather than the stellar rotation period, it provides strong support for a planet-induced origin. The helium triplet at 10830\,\AA, by contrast, is a key tracer of atmospheric escape and is particularly sensitive to radiation-driven interactions in which stellar irradiation affects the extended atmosphere of a close-in planet. 

Interestingly, these lines are located at wavelengths that are challenging for most spectrographs, and that only few spectrographs can observe efficiently. The Ca H \& K doublet is located at the extreme blue of the visible range, in a domain where spectrographs show a much lower efficiency (usually $\sim$1.5-3$\times$ lower than at 550\,nm) due to a combination between optics and CCD efficiency. Since M dwarfs have inherently low flux in the blue, the final result is that Ca H \& K is not an efficient indicators for many of these low-mass host stars. 

On the other hand, the location of He triplet around 1\,$\mu$m is also very challenging. The Si valence bands that allow photoelectron detection impose an upper limit for operation of Si-based detectors (such as CCDs) at 1.2\,$\mu$m and most optical spectrographs choose to stop short of 1\,$\mu$m for efficiency reasons. As such, exception made to a few spectrographs, only near-infrared (NIR) spectrographs are able to measure He triplet emission.

In addition to these magnetic and radiation based interactions, tidal interactions between the star and planet can drive long-term orbital evolution through angular momentum exchange, leading to orbital decay and, in eccentric systems, eccentricity damping. If sufficiently rapid, such evolution may become detectable through secular changes in the radial velocity signal of the system. 

The typical SPI-capable high-resolution spectrographs can be grouped into three sub-types: general-purpose spectrographs, dedicated radial velocity spectrographs, and spectropolarimeters. The discussion below is not intended as an exhaustive census of all facilities relevant to SPI, but rather as a practical overview of the principal classes of high-resolution spectroscopic instruments currently used to probe SPI-related variability, precision time-series signals, and stellar magnetic topology.

\subsubsection{General-purpose high-resolution spectroscopy}

Echelle spectrographs are among the most widely used instrumental designs for high-resolution spectroscopy, as they allow a large wavelength range to be recorded simultaneously while maintaining high spectral resolving power. These instruments use as main dispersive element an echelle grating operated at high diffraction orders, combined with a cross-disperser that spatially separates the overlapping orders on the detector. An additional advantage of these spectrographs is their observational flexibility: several instrumental parameters, such as slit width or cross-disperser configuration, can be adjusted to optimize quantities such as spectral resolution, wavelength coverage, and throughput.

A widely used spectrograph and example of this architecture is UVES, the Ultraviolet and Visual Echelle Spectrograph \citep{2000SPIE.4008..534D}, mounted on one of the VLT Unit Telescopes at Paranal Observatory. UVES is a cross-dispersed echelle spectrograph covering a wavelength range from about 300 to 1100 nm, with a maximum resolving power of approximately $R \sim 80,000$ in the blue arm and $R \sim 110,000$ in the red arm. UVES can cover the full wavelength range of about 300 -- 1060\,nm in two exposures. A closely related instrument in terms of design and operation is HIRES \citep{1994SPIE.2198..362V} on the Keck I telescope at Mauna Kea Observatory. HIRES covers a wavelength range of roughly 300 to 1000\,nm, again in two separate exposures, and can reach a resolving power of about $R \sim 85,000$ when used with its narrowest slit. In both instruments, the availability of multiple instrumental configurations provides substantial observational flexibility, but this comes at the expense of reduced setup reproducibility, which can limit the precision of repeated measurements, in particular radial velocities. This limitation can be mitigated by inserting an iodine absorption cell into the optical path, allowing instrumental profile variations to be tracked and enabling high-precision radial-velocity measurements.

A different instrumental trade-off is represented by X-shooter \citep{Vernet11}, which was designed to maximize simultaneous wavelength coverage by extending observations from the near-ultraviolet well into the near-infrared in a single exposure. This broad spectral range provides a major scientific advantage for SPI studies, since it allows multiple diagnostics to be monitored simultaneously across a wide wavelength range. The price for this versatility is a more modest spectral resolving power (reaching $R \sim 11,000$) than that of classical high-resolution echelle spectrographs such as UVES or HIRES.

For observations further into the near-infrared, up to the practical limits imposed by telluric transmission, CRIRES \citep{2014Msngr.156....7D,2014SPIE.9147E..19F} and its upgrade CRIRES+ \citep{CRIRES+} provide access at considerably higher spectral resolution. The CRIRES to CRIRES+ upgrade increased the wavelength range covered significantly. The spectral resolution of the instrument is either $R> 100\,000$ or $R=50\,000$ depending on the slit width used and the bands covered are Y,J,H,K,L, and M. In this case, however, the gain in resolving power comes at the expense of a much smaller instantaneous wavelength coverage. As a result, X-shooter and CRIRES+ can be viewed as complementary instruments: the former is optimized for broad simultaneous spectral coverage, whereas the latter is better suited to detailed high-resolution studies of selected near-infrared diagnostics.

\subsubsection{Dedicated Radial-Velocity Spectrographs}

A separate class of high-resolution spectrographs has been developed specifically to deliver precise radial-velocity measurements, of the order of m\,s$^{-1}$ and better. In contrast to the more flexible operation of classical echelle instruments, these spectrographs are designed with highly stable opto-mechanical and thermal architectures, with only a limited set of observing modes, restricted to different fiber feeds or detector readout schemes. By minimizing variations in the instrumental profile and maximizing long-term repeatability, these instruments are not only ideal for exoplanet detection, but also particularly valuable for SPI studies that require precise time-series monitoring of subtle, orbit-phased spectroscopic variability \citep[for more see e.g.][]{2018ASSP...49..181F}.

HARPS, the High Accuracy Radial velocity Planet Searcher \citep{2003Msngr.114...20M} provided the prototypical realization of this design philosophy. HARPS is a fibre-fed high resolution \'echelle spectrograph installed at ESO 3.6\,m telescope in La Silla, Chile. The spectrograph is enclosed in a vacuum vessel to avoid spectral drift due to temperature and/or air pressure variations. The spectral range covered is 378\,nm--691\,nm and the spectral resolution $R=115\,000$. As the first instrument to routinely achieve sub-m\,s$^{-1}$ radial-velocity precision, it transformed precision exoplanet spectroscopy and enabled the detection and confirmation of a large number of low-mass planets.

This design paradigm was further developed with ESPRESSO \citep{Pepe21}, which represents a natural evolution of HARPS. ESPRESSO operates in the optical domain (380–780 nm) with a resolving power up to $R \sim 140,000$, and is designed for extreme radial-velocity stability through a vacuum-enclosed, thermally stabilized opto-mechanical system fed by optical fibers from the VLT Unit Telescopes. Its high throughput and stability enable radial-velocity precision down to the $\sim$10 cm\,s$^{-1}$ level. The coupling with the increased photon-collecting power of the VLT reduces photon-noise contribution to radial-velocity measurements, enabling the detection of Earth-mass planets in the habitable zones of solar-type stars. While this can be seen as a relatively direct extension of the HARPS concept, the technical challenges involved in achieving this level of precision are substantial.

In parallel, alternative strategies have been pursued, motivated in part by the high abundance of M dwarfs in the Galaxy. Owing to their lower masses, these stars exhibit larger reflex velocities for a given planetary companion mass, facilitating the detection of low-mass planets. However, this advantage is offset by their intrinsic faintness at optical wavelengths. Efficient observations of M dwarfs therefore require a shift toward the red optical or near-infrared, where new challenges arise, including the need for different detector technologies (e.g., infrared arrays rather than CCDs) and the stronger impact of telluric absorption. Representative instruments following this approach include MAROON-X, CARMENES, and NIRPS. MAROON-X \citep{Seifahrt18} is a high-resolution optical spectrograph (500–920 nm) with $R \sim 80,000$, installed on the Gemini North telescope and with a red wavelength coverage specifically tailored for high-precision radial-velocity measurements of low-mass stars. It was one of the first spectrographs installed to target these stars, opting for the technical solution of going into the extreme red but remaining short of 1\,$\mu$m. A more traditional example of a near-infrared design is CARMENES \citep{Quirrenbach14}, that consists of two spectrographs operating simultaneously in the optical (520–960 nm, $R \sim 94,000$) and near-infrared (960–1710 nm, $R \sim 80,000$), enabling precise radial-velocity measurements across a broad wavelength range. Its dual-channel design is particularly suited to disentangling wavelength-dependent stellar activity from planetary signals in M dwarfs. NIRPS \citep{2017SPIE10400E..18W} is a near-infrared high-resolution spectrograph (980–1800 nm) with a resolving power of $R \sim 75,000$ in high-resolution mode, designed to operate in tandem with HARPS at the ESO 3.6 m telescope. It employs adaptive optics and fiber-fed stabilization to optimize radial-velocity precision for cool, low-mass stars.

\subsubsection{High resolution Spectropolarimeters}

Spectropolarimetry provides an essential diagnostic for star–planet interaction studies, as it enables the direct characterization of stellar magnetic fields and their large-scale topology. This is particularly relevant for SPI scenarios in which magnetic coupling between the star and the planet plays a central role. High-resolution spectropolarimeters therefore offer a unique observational window by combining detailed spectral information with measurements of polarisation of light, allowing the reconstruction of magnetic field geometries through techniques such as Zeeman–Doppler imaging \citep[e.g., Chapter 1 and][]{1989A&A...225..456S}.

The modern development of high-resolution stellar spectropolarimetry is largely rooted in instruments such as ESPaDOnS \citep{2003ASPC..307...41D} and its counterpart NARVAL \citep[e.g.,][]{NARVALexample}, which established the standard design for optical spectropolarimeters. These fiber-fed, cross-dispersed echelle spectrographs operate over the optical wavelength range ($\sim$370–1000\,nm) with resolving powers of $R \sim 65,000$–80,000, and are equipped with dedicated polarimetric modules enabling full Stokes parameter measurements. Their stability and broad wavelength coverage have made them key instruments for reconstructing stellar magnetic topologies and for studying magnetically driven variability.

Building on this design, more recent instruments have extended high-resolution spectropolarimetry into the near-infrared, where cool stars emit most of their radiation and where additional magnetic diagnostics become accessible. A prominent example is SPIRou \citep{2018haex.bookE.107D}, which covers the full YJHK bands (0.95--2.5\,$\mu$m) in a single exposure, both in unpolarized light and in polarized light (circular and linear), at a resolving power of $R \sim 70,000$. By combining radial-velocity precision with full spectropolarimetric capability, SPIRou is particularly well suited to the study of magnetically active low-mass stars and their planetary systems. The extension into the K band, aimed at leveraging the increased spectropolarimetric signal at longer wavelengths, makes SPIRou a particularly ambitious and technically demanding instrument to design and operate.

In parallel, polarimetric capabilities have also been implemented in spectrographs that are not exclusively dedicated to spectropolarimetry. Instruments such as HARPSpol, CRIRES+, and PEPSI offer optional polarimetric modes in addition to their primary spectroscopic configurations. These systems typically achieve higher spectral resolving power or broader wavelength specialization, but often with limited polarimetric flexibility, lower efficiency, or lower repeatability when compared to dedicated spectropolarimeters.

HARPS\-pol extends the HARPS design into the polarimetric domain at optical wavelengths with $R \sim 115,000$, providing high stability for combined radial-velocity and magnetic-field studies \citep{2008SPIE.7014E..0OS}. The design of the polarimeter took advantage of the two fibers entering the spectrograph. Two dual-beam systems can be positioned in the beam: one with a rotating superachromatic quarter-wave plate for circular polarimetry and one with a rotating superachromatic half-wave plate for linear polarimetry. CRIRES+ offers high-resolution ($R \sim 100,000$) near-infrared spectroscopy with polarimetric capability over a narrower instantaneous wavelength range, enabling detailed studies of selected diagnostics up to 5\,$\mu$m, providing an important differentiating factor \citep{2022SPIE12184E..1FD,CRIRES+}. PEPSI \citep[the Potsdam Echelle Polarimetric and Spectroscopic Instrument,][]{2015AN....336..324S}, mounted on the Large Binocular Telescope, reaches very high spectral resolution ($R \sim 130,000-250,000$) and includes polarimetric modes, making it particularly powerful for detailed line-profile and magnetic analyses. In conjunction with the standard wavelength calibration method, PEPSI radial velocities can be obtained relative to a Fabry-P\'erot etalon reference.

\subsection{Low and intermediate resolution spectroscopy}

Most of the ground-based spectroscopic work on SPI is done using high resolution spectrographs. However, low-resolution spectrocoscopy is very useful to study phenomena that broad spectral coverage, or when high instrument throughput and temporal cadence are more important than resolving very fine spectral details. An example of these phenomena are coronal mass ejections (CME), that potentially can cause erosion of planetary atmospheres, and are often studied using lower resolution spectrographs. 

From the ground, stellar CMEs are usually studied using time-resolved neutral-hydrogen H$\alpha$-line spectra that shows radiation from cool plasma of a few times 10,000\,K. The velocities measured in CMEs and also the velocity dispersion of the material is so high that usually low and medium resolution time-series observations of H$\alpha$ line are sufficient. Occasionally, high-resolution spectrographs have also been used for CME studies, e.g., the ESPaDOnS observations of M dwarf V374\,Peg that resulted in a clear detection of CME signature \citep{Vida2016} and Bohyunsan optical echelle spectrograph \citep[BOES,][]{BOES} observations that were a part of multi-wavelength, multi-instrument CME study of a young solar analogue EK\,Dra \citep{Namekata+2026}. 

There are two main philosophies for studying stellar CMEs with low and medium resolution spectrographs: using multi-object spectrographs to study several stars in a cluster at one go, or concentrating on observing interesting, usually highly active, targets one at a time using a long-slit spectrograph. Most of these low-resolution spectrographs have spectral resolving power of 2,000--5,000, meaning that they cannot detect slow CMEs with small velocity dispersion. At times, intermediate resolution observations with 10,000--30,000 have also been used, e.g., ARCES \citep{ARCES} on the ARC 3.5 m telescope at Apache Point Observatory resolution and CHIRON \citep{CHIRON} at the SMARTS 1.5 m telescope at CTIO.

The multi-object instruments that have been used to monitoring young open clusters in hopes of catching flares and associated CME include, EFOSC2 \citep{EFOSC2} at ESO's New Technology Telescope, OSMOS \citep{OSMOS} at the Hiltner Telescope at the MDM Observatory, and ALFOSC at the Nordic Optical Telescope. All these instruments use slit masks that are manufactured separately for each cluster/pointing based on the imaging carried out before the main observing run. The strength of this approach is that one can monitor more stars at one go and that these stars as cluster members are also coeval. When using several clusters of different ages, this would also allow studying the evolution of CME rate with age. However, the clusters that fit completely into a typical field-of-view of such an instrument (some arcminutes) are distant and hence faint. On the other hand, close-by clusters with brighter members tend to have much larger extent than the instrument field-of-view. This means that in practice it is difficult to have more than a few dozens of stars monitored at once. 

One can also use multi-object observations designed to different scientific objectives. One example of serendipitous CME searches is the LAMOST Medium-Resolution Spectroscopic Survey \citep{LAMOST} that observed 226,194 late-type main-sequence stars. \cite{Lu+2022} reported three stellar CME candidates on three M dwarfs in the data sample of more than 1,3 million observations. 

When concentrating in observing one target at the time usually low-resolution long-slit spectrographs are used. Some of the instruments that have been used for this work include EFOSC2 and ALFOSC, that have also been used for multi-object work, and other spectrographs like KOOLS-IFU \citep{KOOLS-IFU} at the 3.8 m Seimei telescope at Okayama Observatory \citep{Seimei}, MALLS \citep{MALLS} on the 2.0-m Nayuta telescope, and BFOSC at NAOC 2.16\,m telescope. In general, concentrating in one target naturally decreases the chances of observing CMEs, as even in the most active stars the large, detectable events are rare.

\section{Ground-based optical and near-infrared photometry}

Ground-based optical and near-infrared (NIR) photometry provides a complementary perspective on star--planet interactions to that obtained from spectroscopy and spectropolarimetry. 
While spectroscopic techniques probe localized diagnostics of stellar activity, magnetic fields, and atmospheric structure, photometry measures the integrated response of the stellar photosphere and chromosphere, capturing variability associated with surface inhomogeneities, transient energetic events, and phase-dependent brightness modulation, including both stellar rotation phase and planetary orbital phase \citep{Shkolnik18}. 
This means photometry is sensitive to a different, but overlapping, set of SPI-relevant observables and offers an essential context for interpreting spectroscopic measurements \citep{Walker08}.

Many SPI signatures, such as rotational modulation and variability correlated with the rotation of a companion, manifest as low-amplitude brightness variations which are approximately coherent over multiple stellar rotations or planet orbits \citep{Shkolnik05, Cauley19}. 
Photometric instruments are therefore effective for SPI when they are designed or operated to deliver highly stable time series over long baselines, enabling direct comparison with contemporaneous spectroscopic and spectropolarimetric diagnostics. Still, fast-evolving features, such as stellar flares, require high photometric precision on timescales of minutes. However, these remain challenging for stars other than the sun, and only the largest flares are detectable \citep{Davenport16}.

From an instrumentation perspective, SPI remains a technically demanding application of ground-based photometry. 
Many SPI-related signals occur at amplitudes of millimagnitudes or below and on timescales that overlap with atmospheric and instrumental noise sources.
As a result, the suitability of a given photometric instrument for SPI studies depends on its stability, calibration strategy, and observing architecture, rather than on its raw sensitivity alone.

\subsection{Fundamental instrumental challenges}
A significant limitation of ground-based photometry arises from the Earth’s atmosphere. 
Atmospheric scintillation introduces high-frequency noise that scales with telescope aperture, airmass, and exposure time, setting a practical noise floor even for bright targets \citep{Young67}. 
On longer timescales, variations in atmospheric transparency, seeing, and extinction introduce correlated noise that can overwhelm signals of stellar rotation or on planetary orbital timescales \citep{Pont06}.

Instrumental effects further complicate SPI observations. 
Detector non-linearity, flat-field uncertainties, small changes in the pointing of the telescope, and inter- and intra-pixel point-spread function variations can all imprint spurious variability signals on both ground-based and space-based photometry.
These effects are particularly relevant for SPI studies, where the signal may be slowly-varying and measured over many nights or months and therefore sensitive to subtle instrumental instabilities.

Commonly, atmospheric effects are mitigated by comparing a star's brightness to a population of nearby stable reference stars \citep{Honeycutt92}, but the effectiveness of this approach depends on there being a sufficient number of nearby stars of similar magnitude and the stability of the detector on pixel and sub-pixel levels.
In practice, the achievable precision is often limited by the stability of the instrument and its environment rather than photon noise

\subsection{Photometric precision from the ground}

Under favourable conditions, ground-based optical photometry routinely achieves millimagnitude precision, with sub-millimagnitude performance possible for bright targets using stable instruments and carefully designed observing strategies. This regime is sufficient to detect rotational modulation from starspots, large flares, and some orbit-phase-dependent variability relevant to SPI. 
For SPI studies, generally a key requirement is not absolute precision achievable in a single snapshot, but the ability of an instrument to deliver consistent, repeatable photometry over long time baselines with well-characterised systematics.

Achieving performance from the ground relies on instrumental approaches that prioritise photometric stability over angular resolution. 
In particular, the use of engineered diffusers to deliberately broaden and stabilise the point-spread function has become an effective technique for mitigating guiding errors, seeing variations, and intra-pixel sensitivity effects. 
By distributing stellar flux over many pixels in a highly repeatable manner, diffuser-assisted photometry reduces sensitivity to atmospheric and instrumental fluctuations and enables near space-like precision for bright stars. 
Purpose-built diffuser-equipped systems, such as the diffuser-assisted photometry mode demonstrated on the ARCTIC imager at Apache Point Observatory \citep{Stefansson17}, have achieved sub-millimagnitude precision in broadband optical filters on timescales relevant to SPI monitoring.

Additional instrumental factors contributing to high-precision performance include detectors with excellent linearity and low read noise, stable thermal and mechanical environments, and guiding systems capable of maintaining consistent illumination patterns across a night or between nights. 
When combined with careful calibration methodology, these design choices allow modern ground-based photometric instruments to detect SPI-related signals.

In the NIR, photometric precision is generally lower due to increased thermal background, stronger telluric absorption, and less well characterised detector systematics. 
However, NIR photometry offers complementary diagnostic power: the reduced contrast of cool starspots relative to the optical can help disentangle magnetic activity from other variability sources, and NIR observations can generally probe different layers of stellar atmospheres, with less chromospheric contribution \citep{Berdyugina05}.
NIR-capable instruments that have been applied to monitoring exoplanet host stars in this context include the TRAPPIST facilities \citep{Jehin11}, SPECULOOS \citep{Delrez18}, and the PINES survey on the Perkins Telescope Observatory \citep{Tamburo22}, which have been used predominantly for M, L, and T dwarf hosts where the NIR contrast advantage is most pronounced.
Combining observations across multiple bandpasses observed simultaneously can therefore provide powerful leverage to break degeneracies between possible stellar surface effects when reducing a 2-d stellar surface to a 1-d time series.

\subsection{Instrument classes for ground-based photometry}

Ground-based photometric instruments relevant to SPI studies can be broadly grouped into several classes.
Each class of instrument provides different trade-offs between field of view, cadence, stability, and flexibility, occupying a distinct niche in SPI analysis.

\paragraph{Narrow-field imagers.}

Narrow-field imagers mounted on small- to medium-aperture telescopes are effective tools for dedicated, detailed monitoring of individual systems.
Narrow-field instruments prioritise stability and repeatability, typically applying a differential photometry approach where comparison stars in the same field are used to remove atmospheric trends. 
In an SPI context, these are typically less useful for untargeted searches\footnote{Although in the \textit{TESS} era, with time-series photometry freely available for essentially all bright stars to inform target selection, truly untargeted pencil-beam searches are less common.}, but are well-suited to long-term monitoring of known planet hosts, particularly when the goal is to measure rotational modulation amplitudes, search for orbit-phase-correlated variability, or characterise flare statistics.

The MEarth Project’s arrays of eight 0.4 m robotic telescopes in both the northern hemisphere (at Fred Lawrence Whipple Observatory, Arizona) and the southern hemisphere (at Cerro Tololo Inter-American Observatory, Chile) enable sustained monitoring of nearby M dwarfs across nearly the entire sky, making them particularly valuable for long time-baseline variability studies \citep{Nutzman08}.

Because these systems often operate in relatively fixed instrumental configurations and are optimised for repeatable differential photometry, they are particularly valuable for SPI studies that require consistency across stellar magnetic cycles, comparisons of activity states over time, or monitoring of systems where orbital geometries evolve due to tidal or magnetic interactions.

Ground-based observations can enable simultaneous multi-band photometry, or fast changes between filters. Because spots, faculae, flares, and extinction residuals have distinct wavelength dependencies, observing in multiple bands nearly simultaneously allows the chromatic behaviour of variability to be characterised effectively.

Instruments such as ULTRACAM at NTT/La Silla \citep{Dhillon07} and HiPERCAM at GTC/La Palma \citep{Dhillon21} implement dichroic beam splitters to observe in several optical bands simultaneously at high cadence. The high, sub-second cadence is well suited to resolving flare morphologies and synchronising optical observations with simultaneous radio or X-ray monitoring. 
Simultaneous multi-band capability is especially important when attempting to distinguish between spot-driven rotational modulation and planet-induced magnetic or thermal signatures, as the spectral slope of the variability often carries more diagnostic information than its amplitude alone.

A related class of multi-band instrument is the MuSCAT family of cameras (MuSCAT, MuSCAT2, MuSCAT3, and MuSCAT4), which use dichroics to observe in three or four optical bands simultaneously on dedicated 1-2m telescopes at Okayama (Japan), Teide Observatory (Tenerife), Haleakal\={a} (Maui), and Paranal (Chile).
Their wide geographic distribution enables nearly continuous longitude coverage of target systems across multiple nights, making them well-suited to the multi-band chromatic diagnostics described above \citep{Narita15, Narita19, Narita20}.

\paragraph{Wide-field survey cameras.}
Wide-field survey instruments provide the ability to monitor very large samples of stars with uniform cadence, enabling population-level SPI tests and variability studies, albeit typically at the expense of higher per-target precision enabled by narrow-field imagers. 
These facilities are also effective for capturing rare of impulsive events (such as megaflares), and for establishing long-term photometric baselines which can later be stitched together into higher-precision observations of individual systems of interest. 

Some examples include dedicated transit-survey cameras such as NGTS \citep{Wheatley18}, WASP \citep{Pollacco06}, and HATNet/HATSouth \citep{Bakos04}, which had the primary goal of finding transiting planets. 
However, time-domain survey facilities not designed specifically for exoplanets can also be highly relevant for SPI through their regular high-precision photometry aimed for supernovae detection or near-Earth object onitoring, including ASAS-SN\citep{Shappee14}, ZTF \citep{Bellm19}, and ATLAS \citep{Tonry19}. 
In the near future, Rubin Observatory's Legacy Survey of Space and Time \citep{Ivezic19} and the Schmidt Observatory's Argus Array \citep{Law22} will provide long-term, densely sampled, multi-band photometry, offering a powerful discovery and characterisation dataset for long-term SPI-related variability in both hemispheres.

\paragraph{Robotic telescope networks.}
Networks of robotic telescopes, such as the Las Cumbres Observatory (LCO) global network \citep{Brown13}, offer a unique capability from single-site facilities. 
By coordinating observations across multiple sites spanning a wide longitude range, these networks can provide nearly continuous coverage of a target system, minimising gaps introduced by day--night cycles and adverse weather at any single site. 
This is particularly valuable for SPI studies that require uninterrupted coverage over planetary orbital periods, or for ensuring many consecutive transits can be observed of a single system with similar instrumentation. 
The LCO network has been used for exoplanet host monitoring and can be scheduled dynamically to respond to real-time alerts from other facilities.

\subsection{Calibration, stability, and long-term monitoring}
SPI studies require a high level instrumental stability as the signals can be low in amplitude and occur on timescales of many days or weeks. 
Long-term photometric monitoring requires consistent detector performance, stable flat fields, and any changes in the instrument response to be well-characterised. 
Even small changes in the properties of the detector can overwhelm signals that span months or years. 
As a result, instruments useful for SPI from the ground benefit from stable observing modes, minimal instrument configuration changes, and careful tracking of calibration data, similar to the requirements of precision radial velocity searches for exoplanets. 
Archival value for these datasets is particularly high, with retrospective SPI analyses covering decades or more across these surveys enabled as new planetary systems are discovered.
The availability of all-sky astrometric and photometric catalogues, particularly from Gaia \citep{Gaia}, has improved the identification of photometrically stable comparison stars for differential photometry, enabling more reliable detrending across multi-season baselines.

\subsection{Limitations and realistic expectations}
Ground-based photometry for SPI applications remains fundamentally limited by atmospheric effects and instrumental systematics. 
For many SPI applications, especially those requiring parts-per-million precision or continuous uninterrupted coverage, space-based observations are required.
Nevertheless, ground-based photometry occupies a critical niche by enabling flexible, long-term, and multi-site observations that complement space missions and provide essential context for multi-wavelength SPI studies.

\section{Radio observations}

In this section an overview of radio observational techniques relevant to stellar and exoplanet studies is provided. The focus will be on frequencies below around 30\,GHz where the emission is overwhelmingly non-thermal and traces high-energy processes that are key to understanding stellar magnetic activity and the ensuring atmospheric forcing on exoplanets. 

Although the boundaries are not strict, radio emission from stellar coronae, winds, and star–planet interactions can broadly be divided by frequency regime: 
\begin{itemize}
\item thermal emission from the corona and wind in the centimetre and mm-wave bands ($\nu > 30$\,GHz), 
\item incoherent (gyro-)synchrotron emission from flare accelerated charges in the decimetre-wave band ($1< \nu < 30$\,GHz), and
\item coherent plasma and cyclotron maser emission in the metre-wave band and lower frequencies ($\nu< 1$\,GHz).
\end{itemize}

A key difference between optical and radio techniques is the large photon occupation in the radio bands allows for phase-coherent detection\footnote{Sometimes called heterodyne detection.}. This allows sampling electric field's phase and amplitude, and combine the fields sampled at different spatial locations `after the fact' on a computer.

\subsection{Historical perspective}
Observational stellar radio astronomy can be said to have begun in the early 60's with claims of radio flare detection on UV Ceti types stars \citep{lovell1963,lovell1964}. These observations were motivated by observations of optical flares on stars that were orders of magnitude more luminous than solar flares, promoting the hypothesis that the radio bursts could similarly be orders of magnitude more luminous and, therefore, detectable. The earliest reported detections had flux densities of tens of Janskys\footnote{For comparison, the brightest solar radio burst in ten years of monitoring data of \citet{saint-hilaire_2013} would have only been $\sim 1\,$mJy at a distance of 10\,pc.} \citep{lovell1969}. However, they were made with a single dish radio telescopes which lack angle-of-arrival information leaving open the possibility that there were human-generated interference. With the advent of interferometers and their secure detection of stars at more modest flux densities \citep{davis1978,kundu1987,kundu1988} it was realised that some of the brighter single dish detections could have been mistaken radio frequency interference. Regardless, it was clear from brightness-temperature arguments that the radio flares had to be non-thermal in origin \citep{hjellming1971}. The detected radio emission was probing accelerated charges and not the hot thermal plasma that powered the bulk of the optical and soft X-ray band emission. 

The advent of more sensitivity radio telescopes in the 80's and 90's led to radio detection of a larger population of stars across the stellar colour-magnitude diagram. Notable early examples include detections with the Jansky Very large Array \citep[VLA, ][]{gary1981,kundu1985,kundu1987,bastian1987}, the Arecibo single-dish telescope \citep{spangler1974}, Westerbork Synthesis Radio Telescope \citep[WSRT, ][]{felli1982} the Australia Telescope Compact Array \citep[ATCA, ][]{beasley1992,jones1994} and the Molonglo Radio Telescope \citep{vaughan1987}. These telescopes  provided sufficient spectral and polarimetric information to identify the non-thermal emission mechanisms that were behind the radio flares \citep{dulk1985}. Due to the choice of frequencies of these telescopes ($\sim 1-10$\,GHz) the majority of the detected flares were due to the incoherent gyrosynchrotron mechanism. Coherent stellar emission occurs at the plasma or cyclotron frequency, which for typical stellar coronal densities and magnetic field strengths, falls at somewhat lower frequencies ($\leq 1$\,GHz; though exceptions exist). The gyrosynchrotron emission allowed for constraints on the total energy in the accelerated non-thermal electron population and the ambient magnetic field strength, but they alone could not pinpoint the cause of the particle acceleration. A breakthrough came with the realisation that gyrosynchrotron radio brightness and quiescent X-ray brightness of flare stars and the Sun were empirically related by the so-called Guedel-Benz relationship \citep[][; however see also \citealt{vedantham2022}]{guedel1993, benz1994}. The empirical relationship suggested that flares were the common mechanism responsible for both coronal heating and radio emission in active stars and the Sun. More recently, the availability of sensitive mm-wave instrumentation has led to detections of similar flares \citep{macgregor2018,macgregor2021,burton2022,vargas-gonzales_2023}, although it remains unclear whether these flares are fundamentally different from their lower-frequency counterparts.

New wideband upgrades to then-existing telescopes allowed for the (unexpected) detection of gyrosynchrotron flares from brown dwarfs \citep{berger2001} showing magnetic activity extends to objects of such low masses. Additional detections came of coherent cyclotron maser emission, thought to be of auroral origin \citep{hallinan2006,hallinan2008}, in brown dwarfs as cold as the methane dwarfs \citep{route2012}. Brown dwarfs' quasi-quiescent gyrosynchrotron was found to be much more radio efficient than their stellar counterparts leading to a breakdown of the Gudel-Benz relationship in the lowest mass stars and brown dwarfs \citep[collectively called ultracool dwarfs;][]{williams2014}. More recently, \citet{kao2023} and \citet{climent2023} used very-long-baseline interferometry (VLBI) to show that the quasi-quiescent emission in at least one ultracool dwarf (SpT of M9) has the morphology of a radiation belt rather than a stellar active region, showing that such `planet-like' attributes start to appear at the end of the stellar main sequence. 

In the meter-wave band, historical interferometric observations struggled with non-detection due to a lack of sensitivity. The recent spate of high-sensitivity and large field of view metre-wave interferometers such as the Murchison Widefield Array \cite[MWA; ][]{mwa}, Low Frequency Array \citep[LOFAR; ][]{lofar} and the upgraded Giant Metrewave Radio Telescope \citep[uGMRT; ][]{ugmrt}, have opened up the metre-wave window. Notwithstanding some exceptions, as expected, the metre-wave band is dominated by coherent emission, especially due to the cyclotron maser instability \citep{lynch2017,vedantham2020,callingham2021,mohan2024}.

\subsection{Observing techniques}
\subsubsection{Single-dish telescopes}
Single dish telescopes focus radiation falling on an aperture -- typically, a dish -- onto the focal point. This type of telescope gives a single `pixel' on the sky. The point spread function of the pixel (the beam) has a width that is given by the diffraction limit of the dish --- $\theta_{\rm d} = \Delta\theta \approx \lambda/D$ where $\lambda$ is the wavelength and $D$ is the diametre of the dish. For a feed that uniformly accepts radiation falling on the dish and fully rejects radiation outside the dish aperture, the beam pattern on the sky is given by the Fraunhofer diffraction of a circular aperture:
\begin{equation}
I(\theta) = \left(\frac{J_1(x)}{x}\right)^2,\,\,x = \frac{2\pi}{\lambda}D\sin\theta,
\end{equation} 
where $\theta$ is the angle from the principal axis of a directional antenna's main beam, so-called boresight. The key problem in stellar observations with single dish telescopes is that it lacks direction-of-arrival information: there is no easy way to disinguish between a faint source at boresignt and a brighter source picked up from the sidelobes of the beam. The second issue is that for typical radio wavelengths and practical dish sizes, the width of the beam, which is the angular resolution of the telescope, can span several arcseconds and lead to confusion between multiple sources within the beam.  

As the dish tracks a target at declination of $\delta_t$, the differential sidereal rotation of an off-axis source at declination $\delta_s$ is approximately $15''\left(\cos\delta_t-\cos\delta_t\right)\,{\rm sec}^{-1}$. Since the sidelobes have an angular scale of $\approx \lambda/D$, the time-scale at which off-axis sources' flux varies is approximately given by 
\begin{equation}
\Delta t \approx \frac{c/(fD)}{15''}\frac{1}{|\cos\delta_t-\cos\delta_s|}.
\end{equation}
In other words, temporal changes in flux due to off-axis source must be expected on timescales of $\theta_{\rm d}/15''$\,seconds and above. 

The sidelobes also scale with frequency.
Since the beam response only depends on the product $f\sin\theta$, the beam response varies by the same amount for the same fractional change in frequency or in $\sin\theta$. Since the beam varies on an angular scale given by $\delta\sin\theta \approx c/(Df)$, the spectral scale over which the sidelobe response of an off-axis source at an angular distance of $\theta$ from boresight varies is given by  
\begin{equation}
\Delta f \approx f \times \frac{c/(Df)}{\sin\theta} = \frac{c}{D\sin\theta}
\end{equation}
In other words, sidelobe confusion should be expected on frequency scales given by $f\theta_{\rm d}$ and above. 

This time--frequency variation of the baseline flux because of bright off-axis source has been an impediment in single dish observations of stellar and brown dwarf radio emission. For instance, consider the star--planet interaction signature due to the cyclotron maser emission mechanism. If the cyclotron maser instability (CMI) emission cone has a width of $\theta_{\rm cmi}$ then the temporal width of the anticipated signal is $P_{\rm syn}\theta_{\rm cmi}/(2\pi)$ where $P_{\rm syn}$ is the synodic period. For a characteristic value of $\theta_{\rm cmi} = 1^\circ$ we have a temporal width of $4\,{\rm min} \times (P_{\rm syn}/{\rm day})$. To avoid baseline sidelobe variation on this timescale we need $\theta_{\rm d}>1^\circ \times (P_{\rm syn}/{\rm day})$. This limits the dish diameter to  $D< 57\lambda \times (P_{\rm syn}/{\rm day})^{-1}$ which severely restricts the sensitivity of observations.

The inability to establish a clean baseline on relevant spectro-temporal scales means that single dish observations have been largely successful for two types of observations. First are rotationally modulated CMI emission from fast rotating brown dwarfs. Typical radio emitting brown dwarfs have spin periods of around three hours or less which means that a $1^\circ$ wide CMI beam can cross the sightline in well within a minute. Indeed, CMI from the methane brown dwarfs was discovered by single dish observations with the Arecibo dish \citep{route2012}. The second area of observational success is the millisecond-structure of CMI emission \citep{abada1997,osten2008,zhang2023}. Here one is detecting elementary CMI sources --- maser sites that are each perhaps just 100\,km in size. On Jupiter the most striking manifestation of elementary CMI sources are the so-called s-bursts --- time-frequency drifts caused by the motion of the source along a field line. Similar drifts have been detected on M-dwarfs \citep{zhang2023} and been used to constrain the magnetic geometry of the emitter \citep{zhang2025}, the along-field drift speed of the elementary sources, and density inhomogeneities in the stellar corona that imprint propagation effects \citep{zhang2026}. On Jupiter, s-bursts are thought to be generated due to sub-Alfv\'enic interaction with Io \citep{hess2007}, although it is premature to posit that the same must be true of similar bursts on stars without detecting the smoking-gun modulation at the planetary synodic period. Nevertheless, observing elementary CMI sources could be a viable method for single dish telescopes to detect star-planet interaction and brown dwarf -- planet interactions in the future.

\subsubsection{Beamforming and interferometry}
Interferometers can overcome some issues of single-dish observations. Interferometers consist of multiple {\em primary elements} which could themselves be dishes. The electric field at the primary elements is measured (phase-coherent detection) and summed on a computer after applying appropriate delays --- an operation called `beamforming'. The essential idea is that the delay-plus-summing is recreating what the parabolic shape of the dish does --- ensure that plane parallel rays from boresight arrive at the focus having traveled the same distance such that they are summed in phase from superposition. The power detected post beamforming can be written as
\begin{equation}
    P_{\rm bf} \propto \left|\sum_{i=0}^{N-1} E_i w_i\right|^2
\end{equation}
where the index $i$ runs over the $N$ primary elements, $E_i$ are the complex electric fields and $w_i$ are the complex beamforming weights. The argument of the weights determine the direction in which the beam is formed. The magnitude of the weights can be all equal which would correspond to uniform illumination but  they can also be chosen to `sculpt' the point spread function to have desirable properties. 
The nominal angular resolution of the interferometer is $\theta_{\rm b} \approx \lambda/B$ where $B$ is the maximum baseline (inter-element separation) in the array. The field of view of the interferometer is given by the diffraction limit of the primary antenna elements: $\lambda/D$ where $D$ is the size of the primary element. While interferometers can achieve high angular resolution (by increasing $B$), the telescope aperture of size $B$ is largely unfilled by the primary elements. This sparsely filled aperture leads to much larger side-lobe levels than that of the Airy function. The confusion created by the higher sidelobe levels can be partially mitigated by the fact that beamforming is a digital operation, so multiple beams in different directions can be formed to serve as `control' beams. The statistics of flux variations in the control beams can be used to establish the level of spectro-temporal fluctuations due to off-axis sources (including human generated interference) to detect true signals from the target beam that exceed this level. Although such rejection of sidelobe artifacts is imperfect, the method has been successfully demonstrated \citep{turner2019,turner2021}. It is worth mentioning here that multiple beams can also be created by having an array of feeds at the focus of dish \citep[e.g.,][]{parkes-13} but the number of beams is limited by optical effects whereas digital beamforming can in principle form as many beams as computational resources allow. 

The real strength of interferometers comes from the recognition that the products $E_iE_j^\ast$, called visibilities, can be computed, averaged in time and frequency to reduce data rates, and stored to disk. This allows the beamforming with weights $w_i$ to be performed offline in any direction, thereby retaining the full angle-of-arrival information. Advanced algorithms exist to efficiently `image' the full primary-element field of view while taking into account Earth rotation and the variation of the weights $w_i$ with frequency and the geometry of primary element locations on the Earth --- a process called synthesis imaging. Modern synthesis imaging methods also partially remove the effect of sidelobes via deconvolution technique such as clean. The majority of contemporary stellar and brown dwarfs observations are taken in this `interferometric mode'. The usual methodology involves interferometric imaging and deconvolution to make a long-exposure synthesis image, post-imaging identification of continuum sources in the field, subtraction of model visibilities of the identified sources from the measured visibilities to remove their sidelobe contribution at the target location, and beamforming of the resulting residual visibilities in the direction of the target at different frequency and time bins to form a dynamic spectrum.

As a short side note, the Very Long Baseline Interferometry (VLBI) technique should be mentioned. It has the essential advantages of interferometers but with very long baselines that provide high angular resolutions down to tens of micro-arcseconds. Stellar VLBI observations have been successfully used to  determine the location and structure of the emitting regions and to constrain orbital motion of exoplanets \citep{massi2008,lestrade1993,fobrich2009,benz1998,curiel2020,kao2023,climent2023}. These observations were all in the decimetric and centimetric bands due to their high angular resolution and brightness of stellar emission.

\subsubsection{Multiplexing}
The fields of view of interferometers is naturally large at longer wavelengths for a given primary aperture size. For instance, LOFAR's 30\,m apertures yields a field of view of $\approx 15\,{\rm deg}^2$ at 150\,MHz which contains around 20 stars within 50\,pc regardless of the pointing direction. This allows for untargeted `piggy-backing' on any data when searching for stellar/exoplanetary emission. This same had not been possible given small fields-of-view at decimetre and centrimetre wavelengths but a new development called phased array feeds (e.g., APRERTIF \citealt{apertif}; ASKAP \citealt{askap}), mitigates this shortcoming. Phased array feeds critically sample the electric field on the focal plane using a two-dimensional array of dipole-like antennas. These measurements are beamformed in real time to illuminate the dish surface so as to form a {\em primary beam} in a given direction. Because the beamforming happens post digitization, a large number of beams can be formed at the same time. Each beam, with width given by the diffraction limit of the dish, is an independent primary beam of the interferometer. For example, the ASKAP phased array feeds \citep{askap} have an instantaneous field of view of $30\,{\rm deg}^2$ even though the diffraction-limited field of view of the dishes' aperture (i.e. single beam) is only  $\approx 1\,{\rm deg}^2$. This level of multiplexing has allowed a large number of radio stars to be identified and studied \citep{pritchard2021,pritchard2024}.

While stellar radio emission can be identified in continuum radio images, many signals of interest occur over short timescales (seconds to minutes) and have peculiar time-frequency signatures such as frequency sweeps.  This requires signal identification in dynamic spectra made in the directions of known stars. Although forming dynamic spectra from visibilities after subtracting background continuum sources is a well-established post processing technique, a recent advance is in the sheer scale of such operations  \citep{tasse2026}. This advance is motivated by the fact that large primary fields of view of a new generation of radio telescopes, particularly at low frequencies ($f\leq 300\,{\rm MHz}$), invariably contains several nearby $d\leq 100\,{\rm pc}$ stellar systems. This means that untargeted surveys of stellar systems and brown dwarfs can be conducted with data acquired for nearly any other science case, amassing large time-on-target/s to search for rare events. Such datasets are perfectly suited to search for space weather causing events \citep{vedantham2022}.  For instance, \citet{callingham2025,konijn2025} used the LOFAR Two Metre Sky Survey (LoTSS) data to amass over 120 star--years worth of data on nearby stellar systems to detect the first stellar type-II burst signaling a coronal mass ejection. With the continued increase in instantaneous field of view of radio telescopes afforded by advances in digital computing, we can expect the utility and success of this method to continue in the foreseeable future.

\section{Space-based instrumentation}

Going beyond Earth's atmosphere offers several advantages in comparison to the ground-based observations. Space missions can provide continuous, high-precision, and multi-wavelength monitoring that is difficult or impossible to achieve from the ground. 

One major advantage is access to wavelengths that are absorbed by the Earth's atmosphere, like the ultraviolet and X-ray wavebands. Many important SPI diagnostics lie in these wavelength regions, that are inaccessible from the ground. Ultraviolet observations with Hubble Space Telescope probe atmospheric escape and chromospheric activity, while X-ray missions such as XMM-Newton and Chandra X-ray Observatory trace coronal heating and magnetic reconnection. Space observations also reach significantly higher photometric precision and stability due to being unaffected by atmospheric turbulence or variable transparency, a source of systematic noise in ground-based data that can mask weak SPI signals. 

Another key strength of space missions is the continuous time coverage. Ground-based observations, depending on the exact waveband used, are often interrupted by the day–night cycle, weather, and seasonal object visibility constraints. This limits the observability of transient or periodic SPI phenomena over long timescales. Space telescopes such as Kepler, K2, and TESS can monitor stars continuously for weeks or months, allowing the detection of flare statistics and stellar rotation signatures much more precisely than would be possible from ground-based observations. Other missions such as CoRoT and CHEOPS, working on low-earth orbits, can have much shorter continuous observations due to earth crossings along the line of sight and data gaps due to crossing South Atlantic Anomaly. However, these mission still offer long observing windows of some 150 days on some fields. Space-based observations are also especially valuable for homogeneous large-scale surveys. Missions such as Kepler, TESS, and Gaia provide consistent datasets for thousands to millions of stars, enabling statistical SPI studies that are difficult to achieve with ground-based campaigns. 

However, ground-based facilities remain essential complements to space missions. High-resolution spectroscopy, spectropolarimetry, and radio observations are only possible from the ground due to the instruments technological complexity and provide crucial diagnostics of stellar magnetic fields, chromospheric emission lines, and possible planetary auroral radio signals. In practice, the most powerful SPI studies combine the strengths of both approaches: stable, multiwavelength space observations together with flexible, high-resolution ground-based data.

We expand on space-based instrumentation subtypes, without attempting to provide an exhaustive review of all instruments available.

\subsection{Infrared observations}

Infrared space-based observations are a powerful probe of SPIs by tracing the thermal response and atmospheric structure of strongly irradiated exoplanets. Unlike optical observations, which mainly probe stellar variability, infrared measurements can be sensitive to (exo)planetary atmospheric heating, heat redistribution, circulation, and chemistry driven by stellar radiation and activity.

The Spitzer Space Telescope \citep{Spitzer}, launched in 2003 into Earth-trailing heliocentric orbit, played a pioneering role in this field. The mission was divided into a 5.5 year-long cold mission, where the full wavelength range of 3--180 $\mu$m could be observed, and into a warm phase lasting from 2009 to 2020, where only two wavelength channels 3.6 $\mu$m and 4.5 $\mu$m were available. Using the Infrared Array Camera \citep[IRAC,][]{IRAC} Spitzer enabled some of the first thermal phase-curve observations of hot Jupiters \citep[see, e.g.,][]{Knutson+2012}. These measurements revealed day-night temperature contrasts, atmospheric circulation patterns, and hotspot offsets caused by strong winds redistributing stellar heat. Spitzer’s major strength was its highly stable long-duration space-based photometry, which allowed continuous monitoring over complete planetary orbits. However, instrumental effects such as intra-pixel sensitivity variations \citep{Reach+2005} and under-sampled point-spread-function due to the large pixel scale of 1.2" \citep{IRAC,Ingalls+2016} required careful calibration, particularly when studying low-amplitude variability.

The James Webb Space Telescope \citep{JWST} is now transforming infrared SPI studies with much greater sensitivity, spectral resolution, and wavelength coverage. Its instruments -- NIRCam \citep{NIRCam}, NIRSpec \citep{NIRSpec}, NIRISS \citep{NIRISS}, and MIRI \citep{MIRI} -- cover wavelengths from about 0.6 to 28 $\mu$m, enabling detailed spectroscopy of exoplanet atmospheres. JWST can measure molecular features such as H$_{2}$O, CO$_{2}$, CO, and CH$_{4}$, allowing studies of atmospheric chemistry, circulation, and thermal structure under strong stellar irradiation.

JWST’s 6.5m main mirror provides significantly improved spatial resolution and photometric sensitivity when compared to the previous infrared missions, reducing contamination from nearby sources and enabling studies of smaller planets and fainter targets. At the same time, SPI studies with JWST still require a careful treatment of instrumental systematics. These systematics include spacecraft issues such as pointing accuracy and detector issues such as large cosmic ray impacts, pixel non-linearity, and signal persistence between exposures \citep[see discussions for example in ][]{Rigby+2023}. There is also so-called "brighter-fatter effect", which is a non-linear detector distortion where bright point sources appear physically larger and blurrier than faint sources \citep{BFE}. All these effect need to be carefully calibrated to obtain data at the required precision levels for exoplanet atmosphere and SPI work.

\subsection{Optical observations}

Optical space-based photometry has been particularly important for studying SPI. Missions such as Kepler, TESS, and CoRoT provide long, nearly uninterrupted light curves that enable detailed studies of magnetic activity of the exoplanet host stars, especially concentrating on flares and starspots. These datasets are especially valuable for searching for enhancements of magnetic activity that are synchronised with planetary orbits, potentially tracing magnetic SPI.

CoRoT, launched in 2006 to an Low Earth Orbit, was the first space mission dedicated to high-precision exoplanet transit photometry and stellar seismology \citep{CoRoT}. Its continuous monitoring capabilities over timescales of weeks to months demonstrated the power of uninterrupted space-based photometry for studying stellar variability and detecting transiting planets around active stars \citep[see, e.g., ][]{Alonso+2008}. CoRoT had a relatively small field of view and pixel scale of 2.32", which reduced blending compared to later wide-field space-based surveys, although crowding could still be significant in dense stellar regions.

Kepler \citep{Kepler} represented a major advancement in photometric precision and long-term stability. It was launched in 2009 into a Earth-trailing heliocentric orbit. The mission was designed primarily for exoplanet transit detection, it monitored over 150,000 stars continuously for several years with parts-per-million (ppm) precision. Its continuous, high-cadence observations made it very powerful for studying stellar rotation cycles \citep{McQuillan+2014}, flare statistics \citep{Maehara+2012,Davenport16}, starspot evolution \citep{Basri+2011}, and subtle orbital phase variations potentially associated with SPIs. Kepler data also enabled the first large statistical studies of stellar activity in exoplanet host stars \citep[see, e.g., ][]{McQuillan+2013}. 

Although Kepler provided exceptionally stable long-duration photometry, instrumental systematics related to pointing drifts, thermal variations, and detector effects required careful calibration and detrending \citep[see, e.g.,][]{Jenkins+2010}. Kepler’s pixel scale was also relatively large, approximately 4" per pixel, meaning that contamination from nearby stars could become significant, particularly in crowded fields. This target confusion is especially relevant for flare studies and low-amplitude variability analyses, where signals originating from nearby active stars may be incorrectly attributed to the target star \citep[see discussion in][]{Bryson+2010}.  

Kepler was continuously pointing at the same area of the sky, increasing its observation baseline at the expense of limiting its capability to probe different stellar populations. Kepler's successor mission, K2 \citep{K2}, expanded the original mission's studies to a wider variety of stellar populations along the ecliptic plane, although with shorter observing campaigns and somewhat reduced pointing stability. 

The next pivotal space-based exoplanet mission, TESS \citep{TESS}, was launched in 2018 into a high Earth orbit. TESS differs from Kepler by emphasizing bright nearby stars across almost the entire sky. Its four wide-field cameras simultaneously monitor large areas of the sky with cadences ranging from 20 seconds to 30 minutes. The observing baseline for a single sector is typically shorter than Kepler’s, normally of 27 days, improving when moving towards the ecliptic poles, where the coverage is continuous. This wide-field strategy comes with a pixel scale of 21", making source confusion and flux contamination major challenges, particularly in crowded Galactic fields. In SPI studies, where weak stellar variability or transient flare signatures are often important, careful vetting is required to ensure that detected activity originates from the planet-hosting star itself rather than a nearby contaminating source. Still, TESS’s large target sample and brightness of the targets make it especially valuable for population studies and identifying interesting targets that are suitable for spectroscopic and multi-wavelength follow-up. 

Astrometric and spectrophotometric missions also contribute indirectly to SPI studies. Gaia \citep{Gaia} primarily provides ultra-precise astrometry, parallaxes, and proper motions for more than a billion stars, but it also delivers broad-band photometry, low-resolution spectroscopy in blue and red prism channels, and sparse time-series measurements. Due to providing accurate distances, Gaia is particularly important for deriving accurate stellar radii, luminosities, ages, and orbital architectures, all of which are essential for interpreting SPI signatures. Its homogeneous characterisation of exoplanet host stars enables improved comparisons between active and inactive systems and helps place SPI phenomena into a broader evolutionary context.

Together, these space missions highlight how space-based instrumentation in visible wavelengths combines photometric precision, long-term stability, homogeneous calibration, and uninterrupted monitoring to provide an important approach for studying SPI, while also emphasising the importance of understanding instrumental limitations such as source confusion, drifts caused by satellite pointing, and high cosmic ray fluxes. 

\subsection{Ultraviolet}

Ultraviolet (UV) observations are central for studying SPIs because they probe both the high-energy stellar radiation field and the extended, escaping atmospheres of close-in exoplanets. UV wavelengths trace chromospheric and transition-region emission lines, as well as resonance lines from escaping material, making them sensitive to atmospheric escape and magnetic activity-driven interactions.

The Hubble Space Telescope (HST) has been the key facility for UV SPI studies, primarily through its STIS \citep{STIS} and COS \citep{COS} instruments. STIS enables high spatial and spectral resolution long-slit and echelle spectroscopy, while COS provides extremely high sensitivity in the far-UV, making it particularly effective for faint targets. Together they cover critical diagnostics such as Ly$\alpha$ (1216 {\AA}), C\,II, Si\,III, and Mg\,II lines. These instruments have enabled time-resolved transit spectroscopy that revealed extended hydrogen exospheres around hot Jupiters \citep[see, e.g., ][]{Vidal-Madjar+2003} and warm Neptunes \citep[see, e.g.,][]{Ehrenreich+2015}, often exceeding planetary Roche lobes and directly demonstrating atmospheric escape driven by stellar irradiation \citep[see, e.g., ][]{Vidal-Madjar+2003}.

From an instrumental perspective, UV observations are photon-limited and require space-based platforms due to the absorption caused by Earth's atmosphere. Hubble’s stability and low background are therefore essential. Observationally UV is a challenging regime, even from space. Due to HST being on Low Earth Orbit, one of the major observational challenges is geocoronal Ly$\alpha$ contamination, where the sunlit upper atmosphere of the Earth emits radiation, particularly in the Far-Ultraviolet (FUV) range \citep[see, e.g., ][]{Bohlin+2014}. This emission acts as a bright, variable foreground that needs to be minimised by scheduling the sensitive observations during orbital night. Additional instrumental effects include detector sensitivity degradation over time, and a relatively small collecting area compared to optical facilities. Precise time-series spectroscopy during transits is also demanding, requiring accurate scheduling and long uninterrupted exposures to capture faint absorption signatures of the escaping gas.

The GALEX mission \citep{GALEX} complemented HST by providing wide-field UV imaging in the near- and far-UV. Although lacking spectroscopic capabilities, its large field of view enabled statistical studies of stellar activity across thousands of stars \citep[see, e.g., ][]{Welsh+2007,Shkolnik+2011}, helping to place planet-hosting systems in a broader context of UV radiation environments.

The Colorado Ultraviolet Transit Experiment \citep[CUTE, ][]{CUTE} is an interesting example of a much smaller mission used for SPIs studies in UV. It is a dedicated ultraviolet \textit{CubeSat} mission designed to study atmospheric escape and star–planet interactions in close-in exoplanet systems. Operating in the near-UV (approximately 2550–3300 \AA), CUTE performs time-resolved transit spectroscopy of hot Jupiters to probe extended planetary atmospheres, mass loss, and interactions with the stellar environment \citep{Sreejith+2023}. Its compact spectrograph focuses on diagnostics such as Mg\,II and Fe\,II lines, which are sensitive to escaping material and stellar activity. Although much smaller than major space observatories and limited in its scope, CUTE demonstrates how dedicated small missions can provide long-duration monitoring of SPI-related phenomena and complement larger UV facilities through targeted, high-cadence observations of individual systems.

Overall, UV space instrumentation provides direct access to the energetic processes driving atmospheric escape and magnetic interaction in SPI systems, with Hubble offering detailed spectroscopic diagnostics and GALEX enabling population-level constraints on stellar UV activity.

\subsection{X-rays}

X-ray emission originates in million-degree plasma providing a direct probe of the highest-energy processes involved in SPIs. These wavelengths provide crucial information on stellar coronae, magnetic reconnection events, and the energetic irradiation that drives planetary atmospheric escape. 

The XMM-Newton observatory \citep{XMM_Newton} has been especially important in SPI studies due to its large effective collecting area and moderate spectral resolution, enabling sensitive measurements of coronal emission from relatively faint exoplanet host stars \citep[see, e.g., ][]{Webb+2020}. The European Photon Imaging Camera (EPIC) consists of three advanced X-ray CCD cameras (two Metal Oxide Semi-conductor CCD arrays, one pn-CCD) designed for high-sensitivity imaging and broad-band spectroscopy in the 0.1--15 keV range \citep{EPIC_MOS,EPIC_pn}. EPIC is well suited for SPI studies as it can detect variability, flares, and long-term changes in coronal temperature and emission measure. XMM-Newton’s high throughput allows for time-resolved studies, which are essential for searching for orbital phased variability or enhanced activity potentially linked to planetary interaction \citep[e.g.,][]{Poppenhaeger_Wolk2014}.

The Chandra X-ray Observatory \citep{Chandra} complements this capability with its superior angular resolution, of about 0.5", achieved through its grazing-incidence mirror design. This high spatial resolution is particularly valuable in crowded stellar fields, where it helps disentangle the X-ray emission of the planet-hosting star from nearby companions that could otherwise contaminate the signal. Chandra’s ACIS \citep{ACIS} and HRC \citep{HRC} instruments also provide high-quality imaging spectroscopy and timing capabilities, enabling detailed flare characterisation and coronal structure studies.

From an instrumental perspective, X-ray SPI studies face several key challenges. X-ray photons are intrinsically sparse, requiring long integrations to build statistically meaningful light curves. Both XMM-Newton and Chandra are therefore typically used in pointed observations rather than survey mode, limiting sample sizes compared to optical missions. Background flaring from charged particles, detector dead time, and the need for careful background subtraction are also important factors, particularly when searching for subtle orbitally modulated signals.

Despite these limitations, X-ray observations are uniquely powerful for SPI research because they directly probe the stellar corona and the high-energy radiation environment important for the atmospheric escape and ionisation in close-in planets and very difficult to characterize from other wavelengths. In some systems, tentative evidence has been reported for enhanced X-ray luminosity or phase-dependent variability in stars hosting hot Jupiters \citep[e.g.,][]{Poppenhaeger_Wolk2014}, although separating planetary effects from intrinsic stellar variability remains challenging. Nonetheless, X-ray diagnostics remain essential for constraining the energy input from the host star and for linking magnetic activity to long-term planetary atmospheric evolution.

\section{Solar System observations}

The solar system offers a detailed view of nearby planetary environments where to study the effect of stellar activity. Solar activity drives upper atmospheric variability across terrestrial planets through both short-term transient events and long-term cyclic modulations, with the magnitude and nature of these responses fundamentally shaped by each planet's magnetic field configuration. 
The earliest studies on our own planet showed that Earth's atmosphere responds to solar activity over a range of timescales, which leads to short and long-term changes globally. Current observational evidence demonstrates that short-term atmospheric responses occur across multiple timescales and atmospheric layers. On timescales of hours to days, solar flares and coronal mass ejections (CMEs) trigger geomagnetic storms and ionospheric disturbances, with electromagnetic radiation driving immediate ionospheric responses and CME impacts on the magnetosphere causing severe geomagnetic storms that propagate effects from the magnetosphere to Earth's surface \citep[for reviews see][]{Tsuda2015, 10.3389/fspas.2023.1244402}. Solar energetic particle events produce dramatic atmospheric effects, with observations showing ozone decreases of 20–40\% during solar proton events accompanied by simultaneous cooling of in the lower mesosphere \citep{https://doi.org/10.1029/2007RG000236}. 

Satellite observations from the Solar Mesosphere Explorer showed that on the 27-day solar rotation timescale, atmospheric temperatures varied by 1.5 K at 2 mbar, growing to 2.5 K at 70 km altitude. Space-based measurements have documented total solar irradiance variations of 0.5\% when large sunspots cross the solar disk and 0.2\% from one month to the next \citep{TSIROPOULA2003469}. Ground-based observations further confirm short-term responses, with Relative Ionospheric Opacity Meter (riometer) measurements detecting polar cap absorption events during major solar storms \citep{2013JSWSC...3A..06K}. Long-term atmospheric responses to solar activity are also evident across the 11-year solar cycle and longer timescales. 

At Mars, where only localized crustal magnetic fields exist, MAVEN observations reveal dramatic short-term responses to solar forcing \citep{https://doi.org/10.1029/2018GL077731,https://doi.org/10.1029/2020JA028518,2025Univ...11..245H}. During interplanetary coronal mass ejection (ICME) impacts, MAVEN STATIC and SWEA instruments detected enhanced suprathermal electron fluxes that drove dayside electron-impact ionization rates to exceed photoionization, producing unusually high minor ion densities and ion escape rates \citep{2017P&SS..145...28D,https://doi.org/10.1029/2024GL111676}. The 27-day solar rotation signal is prominently observed in MAVEN NGIMS and EUVM data, with strongest density variations at 200–250 km altitudes \citep{https://doi.org/10.1029/2021JE007036}. 
At Venus, lacking any intrinsic magnetic field, the induced magnetosphere responds differently. Venus Express ASPERA-4 IMA and ELS measurements during ICME passages showed magnetic barrier strengths exceeding 250 nT (compared to typical 30–40 nT), ionopause compression, and O+ escape fluxes reaching two to three orders of magnitude above background rates \citep{https://doi.org/10.1029/2017JA024852,Xu_2019,https://doi.org/10.1029/2024JA032553}. During corotating interaction regions (CIRs) and CMEs, ASPERA-4 detected ion escape rate enhancements by factors of 2–10, with the unmagnetized nature of Venus allowing direct solar wind erosion of the ionosphere \citep{2008JGRE..113.0B04L,2011JGRA..116.9308E}. 

Ground-based spectroscopic and imaging observations of sodium exospheres \citep{1985Sci...229..651P,https://doi.org/10.1029/GL015i013p01515, 1998JGR...103.8581P,2019Icar..328..152K} around airless bodies have employed progressively higher-resolution instrumentation to resolve the faint Na D-line emission against scattered sunlight and sky background. For the Moon, dual-etalon Fabry-Perot spectrometers mounted on the National Solar Observatory McMath-Pierce Solar Telescope achieved resolving powers of R $\approx$ 180,000, with a field of view of approximately 3 arcminutes ($\approx$ 336 km at lunar distance), enabling limb measurements extending to $\sim$950 km altitude \citep{https://doi.org/10.1002/2014JA019801,https://doi.org/10.1029/2018JE005717}. Complementary coronagraphic observations using the Lyot coronagraph coupled with the DARRK grating spectrograph at Mount Lemmon provided lower spectral resolution (R $\approx$ 5,000) but superior stray-light suppression to within a few arcseconds of the lunar limb \citep{D_M_Hunten_1997}, while narrow-band high-resolution imaging techniques have achieved sensitivity down to tangent altitudes of $\approx$50 km \citep{1995AJ....109..835S}.  
Lunar observations have shown a clear correlation of exospheric sodium line intensity with solar UV and EUV flux \citep{10.1093/mnras/staf1447}. Variations in the lunar exosphere has also been reported during ICME passages \citep{2012JGRE..117.0K02K,2025GeoRL..5215737D}.

Mercury's sodium exosphere has been characterized using even higher spectral resolution: the THEMIS solar telescope equipped with the MTR high-resolution spectrograph reached R $\approx$ 370,000 during daytime observations \citep{2009GeoRL..36.7201L}; the Rapid Imaging Planetary Spectrograph (RIPS) achieved resolving powers up to R $\approx$ 127,000 with simultaneous narrowband slit-viewer imaging; and the Extreme Precision Spectrometer (EXPRES) on the 4.3-m Lowell Discovery Telescope operated at R $\approx$ 150,000 for high-stability line-profile measurements. 

Space missions have provided insitu measurements that have revolutionized our understanding of planetary exospheres by providing direct measurements of neutral and ionized species, their spatial distributions, temporal variability, and responses to external drivers. The primary instruments for exospheric observations are neutral mass spectrometers such as on LADEE \citep{Mahaffy2014}, Chandrayaan-2 \citep{2021GeoRL..4894970D}, MAVEN \citep{2015SSRv..195...49M} and UV spectrometers on MAVEN \citep{2015SSRv..195...75M}, LADEE \citep{Colaprete2015} and Venus Express \citep{BERTAUX20071673} . 

X-ray emission has been detected from all planetary bodies and most moons \citep{2002ApJ...572.1077E} of the Solar System \citep{2007P&SS...55.1135B, https://doi.org/10.1029/2017JA024852,https://doi.org/10.1002/asna.20210101} by either one or all of the following processes.
\begin{itemize}
    \item X-ray fluorescence and scattering is ubiquitous with the intensity varying as a function of the solar flare strength and spectral profile. On planetary orbiters, these measurements so far have been non-imaging and aimed at mapping elemental abundances \citep{1972Sci...175..436A,SWINYARD2009744,2011Icar..214...53N, 2012P&SS...60..217W,2016RAA....16....4D, 2020Icar..34513716N,2024Icar..41015898N}. XRF emission and scattering have also been measured from  the outer planets and Pluto as well as from asteroids.  Jupiter and Saturn’s X-rays come from XRF and scattering from the upper atmosphere . Uranus has a low‑significance X‑ray signal exceeding pure solar scattering expectations, and Neptune shows only upper limits.
    \item Particle induced X-ray emission (PIXE) often episodic have been observed on moons of Jupiter, Mercury and Moon \citep{2012JGRE..117.0K02K}
   \item Jupiter shows clear polar X‑ray aurorae dominated by line emission from highly ionized heavy ions and variable bremsstrahlung. The auroral spectrum is dominated by charge exchange of precipitating highly ionized heavy ions (notably oxygen and sulfur) that produce line emission, with episodes where high‑energy electron bremsstrahlung contributes at $>$ 2 keV during strong events. No convincing auroral X‑ray signature has been detected with current Earth‑orbit observatories and any Saturnian X‑ray aurora is likely below their sensitivity thresholds. Observations are sparse for the ice giants and interpretation of source processes limited by signal levels. The current understanding of X-ray emission influence of solar activity is brought about by high spatial resolution and high spectral throughput instruments on X-ray observatories such as XMM-Newton, Chandra and NuSTAR.
   Earth's X-ray aurorae has been primarily mapped by the PIXIE \citep{1995SSRv...71..385I} instrument on POLAR satellite.  However there are no spectroscopic measurements in the soft X-rays to de-convolve the source mechanisms and their relative contributions. 
   \item Solar wind charge exchange (SWCX) \citep{2010SSRv..157...57D} occurs when highly charged solar wind ions—such as O${^7+}$, C$^{6+}$, Ne$^{8+}$, and heavier species capture electrons from neutral exospheric or atmospheric atoms, producing characteristic soft X-ray emission in the 0.1–2 keV energy range. SWCX serves as a simultaneous diagnostic of both solar wind heavy ion composition and neutral exosphere density, making it a critical tool for understanding atmospheric escape processes and solar wind–magnetosphere coupling. 
   
\end{itemize}

These contrasting responses from solar system planets to solar wind and solar energetic events,  underscore how magnetic field topology, planetary atmospheric density and solar energetic event propagation in the heliosphere, fundamentally modulates solar-driven energy deposition, atmospheric expansion, and ion escape processes. Observation of terrestrial planetary atmospheres in the Solar System provides critical constraints for understanding how stellar activity shapes atmospheric evolution across diverse planetary environments, with direct implications for assessing habitability and atmospheric retention around exoplanets orbiting active stars.

\section{Outlook on future instrumentation}

SPI instrumentation ideally combines high sensitivity, temporal stability, broad wavelength coverage, and flexible observing modes, since SPI signatures are often weak, time-variable, and span multiple wavelength regimes. Continuous high-cadence observations are particularly important for detecting flares, magnetic reconnection events, transit asymmetries, and orbitally phased variability.

High spectral resolution is essential for resolving line-profile variations, chromospheric activity indicators, atmospheric escape signatures, and precise radial velocity shifts, while spectropolarimetry provides direct measurements of stellar magnetic fields that are central to many SPI processes. Together, these capabilities are critical for characterising the magnetic and dynamical interactions between stars and planets.

UV and X-ray instruments are well equipped to probe stellar magnetic activity, coronal heating, and exoplanetary atmospheric escape, while optical and infrared facilities trace stellar variability, exoplanet atmospheres, and thermal phase curves. Low-frequency radio observations are especially important because they may provide the most direct probes of planetary magnetic fields through auroral emission. Ideally, future SPI studies would combine simultaneous observations across these wavelength regimes.

Looking at the future ground-based flagship facilities coming on-line the coming years, there are the 30-meter class telescope in optical and infra-red wavelengths where the European Southern Observatory's Extremely Large Telescope (ELT) is on an advanced stage of construction. In radio domain one can identify the Square Kilometer Array (SKA) and next generation Very Large Array (ngVLA). In the coming years we will also have several important space missions, e.g., Ariel, PLATO and Habitable Worlds Observatory (HWO) working in the visible range, and New Athena in X-rays\footnote{https://www.the-athena-x-ray-observatory.eu/en/newathena-mission}. 

On the ground ELT\footnote{https://elt.eso.org/} is the next transformative project in visible and infra-red wavelengths. From the planned ELT instruments, the high resolution spectrograph ANDES \citep{ANDES2022,ANDES2024}, is the most interesting for SPI studies. ANDES will enable a wide range of transformative studies in exoplanet and stellar astrophysics. For giant exoplanets, it will provide detailed atmospheric characterisation, including constraints on dynamics and weather patterns, while for rocky exoplanets it will probe atmospheric composition and potential biosignatures \citep{ANDES2024}. In parallel, it will deliver high-resolution spectroscopy of stellar activity and magnetic fields, including measurements based on Zeeman broadening and splitting, taking advantage of the ELT’s large collecting area and broad wavelength coverage. This will also enable systematic studies of how stellar mass, age, and environment shape stellar activity and magnetism across different stellar populations. However, a key limitation is that the optical layout of ELT is not well suited for polarimetric observations, and therefore in the planned configuration ELT cannot easily directly measure full magnetic field geometries. In general, relatively few new spectropolarimetric instruments are currently planned, particularly for large and medium-sized night-time telescopes.

Historically, radio observations mainly focused on stellar coronal activity, providing indirect insight into star–planet interactions (SPI). More recently, attention has shifted toward direct SPI signatures, including coronal mass ejections, energetic particle events, and radio emission from brown dwarfs and potentially giant exoplanets. This progress is driven largely by major gains in sensitivity expected from upcoming facilities such as SKA\footnote{https://www.skao.int/en}, ngVLA\footnote{https://ngvla.nrao.edu/}, the Deep Synoptic Array\footnote{https://www.deepsynoptic.org/}, and the FAST Core Array \citep{FAST}. These instruments may enable the first detections of incoherent radiation belt emission or coherent cyclotron maser emission from substellar and planetary-mass objects, opening a new observational window onto magnetic field strengths and radiation environments of massive exoplanets.

In space, PLATO \citep{PLATO} will deliver long-term, high-precision photometry for bright stars, improving the study of stellar activity cycles and tidal interactions, while missions such as ARIEL \citep{Ariel} will focus on comparative exoplanet atmospheres and may reveal how stellar magnetic environments shape atmospheric evolution. Whereas, HWO's\footnote{https://habitableworldsobservatory.org} ability to perform long-term, high-precision photometry and spectroscopy will enable detailed characterisation of stellar variability and potential orbitally modulated signals, helping to disentangle intrinsic stellar activity from subtle planet-induced effects and linking stellar magnetic behaviour to planetary atmospheric response. Another interesting proposed space mission is LIFE \citep{LIFE}, which aims to characterise atmospheres of a statistically relevant sample of small exoplanets.

In the Solar System, planetary missions provide detailed local measurements using orbiters, landers, and rovers, while global-scale observations are essential for connecting these studies to exoplanetary systems. Such global observations have been carried out for Earth\citep{2014ApJ...787..171R,Kelkar_2025,2025ApJ...983..168G,nandi2025shapespectropolarimeteronboard} while there are limited studies for Mars \citep{10.1093/mnras/stab225,2024AGUFMP43C.3027S}and Venus \citep{2025AAS...24523201E}. Particularly promising are wide-field X-ray imaging spectroscopy missions such as SMILE \citep{2022hxga.book...95B}, which can study Sun–planet interactions across entire magnetospheres. Similar approaches from elliptical or aerostationary orbits, as well as future UV/X-ray monitoring from the Moon, could provide continuous observations of planetary responses to solar activity.

In general, the development of future instrumentation is moving SPI research from sparse, indirect detections toward a time-resolved, multi-wavelength and physically diagnostics, where stellar magnetic activity and planetary responses can be studied as coupled systems rather than isolated phenomena. In summary, next-generation instrumentation will enable a shift from isolated detections toward a systematic and physical understanding of SPI processes across diverse stellar and planetary environments.

\section*{Acknowledgements}
PF acknowledges financial support from the Severo Ochoa grant CEX2021-001131-S funded by MCIN/ AEI/10.13039/501100011033. PF is also funded by the European Union (ERC, THIRSTEE, 101164189). Views and opinions expressed are however those of the author(s) only and do not necessarily reflect those of the European Union or the European Research Council. Neither the European Union nor the granting authority can be held responsible for them.

\section*{Conflict of Interest}
The authors have no relevant financial or non-financial interests to disclose, and no competing interests to declare that are relevant to the content of this article.\\

\bibliographystyle{plainnat}
\bibliography{Chapter5}

\end{document}